\documentclass{JHEP3}
\usepackage{graphicx,epsf,epsfig}
\usepackage{amssymb}
\usepackage[square,comma,numbers,sort&compress]{natbib}
\def\eq#1{{eq.~(\ref{#1})}}

\def\vev#1{\left\langle #1\right\rangle}

\def\hbar{\hspace{0pt}\raisebox{1pt}{$-$} \hspace{-7pt} h}

\newcommand{\be}{\begin{equation}}
\newcommand{\ee}{\end{equation}}
\newcommand{\bd}{\begin{displaymath}}
\newcommand{\ed}{\end{displaymath}}
\newcommand{\bea}{\begin{eqnarray}}
\newcommand{\eea}{\end{eqnarray}}
\newcommand{\nn}{\nonumber}

\def\so10{$SO(10)$}

\title{$A_4$ family symmetry and quark-lepton unification}
\date{August 25, 2006}
\author{Stephen F. King\\ School of Physics and Astronomy, University of Southampton, SO16 1BJ Southampton, United Kingdom\\ E-mail: \email{sfk@hep.phys.soton.ac.uk}}
\author{Michal Malinsk\'{y}\\ School of Physics
and Astronomy, University of Southampton, SO16 1BJ Southampton,
United Kingdom\\ E-mail: \email{malinsky@phys.soton.ac.uk}}
\abstract{ We present a model of quark and lepton masses
and mixings based on $A_4$ family symmetry, 
a discrete subgroup of an $SO(3)$ flavour
symmetry, together with Pati-Salam unification.
It accommodates tri-bimaximal neutrino mixing via constrained sequential dominance with a particularly simple vacuum alignment mechanism emerging through the
effective D-term contributions to the scalar potential. }
\begin{document}
\section{Introduction}
There has recently been considerable interest in the use of the
discrete group $A_4$ as a family symmetry
\cite{Ma:2001dn,Ma:2002yp,Babu:2002dz,Hirsch:2003xx,Hirsch:2003dr,Ma:2004pt,Ma:2004zv,Ma:2004zd,Chen:2005jm,Ma:2005sh,Babu:2005se,Ma:2005mw,Ma:2005qf,Zee:2005ut,Adhikary:2006wi,Ma:2006sk,Ma:2006dn,Altarelli:2005yp,Altarelli:2005yx,deMedeirosVarzielas:2005qg,Lavoura:2006hb,Adhikary:2006jx,He:2006dk,Altarelli:2006kf,Altarelli:2006kg}.
A particularly attractive feature of $A_4$ is the possibility of 
obtaining non-trivial vacuum alignment in a simpler way
than for continuos family symmetries
\cite{Altarelli:2005yp,Altarelli:2005yx,deMedeirosVarzielas:2005qg}.
Such non-trivial vacuum alignments are of interest 
since they can lead to tri-bimaximal neutrino mixing
\cite{Harrison:2002er}. In particular, in the framework of the
see-saw mechanism with sequential dominance (SD)
\cite{King:1998jw,King:1999cm,King:1999mb,King:2002nf},
such non-trivial vacuum alignment can lead to constrained
sequential dominance (CSD) \cite{King:2005bj} in which
tri-bimaximal neutrino mixing arises from simple relations
between Yukawa couplings involving the dominant and
leading sub-dominant right-handed neutrinos.

Despite the great interest in $A_4$ as a family symmetry,
there does not yet exist in the literature a model in which 
quarks and leptons are unified. Part of the reason for this
is that the left and right handed chiral components of the
quarks and leptons are usually required to transform differently
under the $A_4$ family symmetry
\cite{Ma:2001dn,Ma:2002yp,Babu:2002dz,Ma:2004zv,Chen:2005jm,Adhikary:2006wi,Ma:2006sk,Ma:2006dn,Altarelli:2005yp,Altarelli:2005yx,deMedeirosVarzielas:2005qg,Lavoura:2006hb,Adhikary:2006jx}.
If both helicity components transform in the same way
then the $A_4$ family symmetry does not prevent trivial invariant
operators which give a mass matrix contribution proportional to the
unit matrix \cite{Ma:2006wm}, rather than the desired hierarchical
form. The situation is rather similar to the case of $SO(3)$ family
symmetry since $A_4$ may be regarded as a discrete subgroup of
$SO(3)$. In the case of $SO(3)$ the solution to this problem is
to accept the left-right asymmetry, and to construct partially
unified models based on Pati-Salam gauge group \cite{King:2005bj}.
Such models can in principle be embedded directly into 
string theory, and may be consistent with $SO(10)$ in a 5D framework
\cite{King:2006me}, without the need for an explicit 4D $SO(10)$
GUT, which in any case suffers from the doublet-triplet splitting problem.
However, to best of our knowledge, no such Pati-Salam unified model with
$A_4$ family symmetry exists in the literature.

In this paper we present a realistic model of quark and lepton
masses and mixings based on $A_4$ family symmetry and Pati-Salam
unification. The model goes along the lines of the $SO(3)$ 
and Pati-Salam model discussed
in detail in \cite{King:2006me}, and shares many of the desirable
features of that model, in particular the flavons entered at the
lowest possible order, which allowed the messenger sector to 
be explicitly specified. Also, as in \cite{King:2006me},
tri-bimaximal neutrino mixing emerges from the see-saw mechanism with
CSD arising from vacuum alignment. However, whereas the vacuum
alignment in $SO(3)$ \cite{King:2005bj}, assumed in \cite{King:2006me},
was rather involved, here, with the discrete
subgroup $A_4$, it will become remarkably simple.
Here we will use the discrete radiative vacuum alignment mechanism
proposed in \cite{deMedeirosVarzielas:2006fc}
for the $\Delta (27)$ discrete symmetry model,
based on discrete D-terms rather than the F-term mechanism
discussed in \cite{deMedeirosVarzielas:2005qg} for discrete
subgroups of $SO(3)$ and $SU(3)$. In fact the $A_4$ model
presented here as a discrete version of the $SO(3)$ models discussed in
\cite{King:2005bj,King:2006me},
mirrors the $\Delta (27)$ model discussed in
\cite{deMedeirosVarzielas:2006fc} which is a discrete version
of the $SU(3)$ models discussed in  
\cite{King:2001uz,King:2003rf,deMedeirosVarzielas:2005ax}.

\section{The model}
The model is based on a high-energy Pati-Salam $SU(4)_{C}\otimes
SU(2)_{L}\otimes SU(2)_{R}$ supersymmetric model with Yukawa sector
driven by a discrete subgroup of $SO(3)$, the $A_4$ flavour symmetry
and a pair of extra symmetry factors $U(1)\otimes Z_{2}$ to forbid
some unwanted operators.
The construction goes along similar lines as in the case of a fully
$SO(3)$ invariant model studied in \cite{King:2006me}. However,
sticking to a discrete subgroup of a Lie-group brings in several
qualitative changes that require a separate treatment. In particular,
it provides for a very effective tool to address the vacuum alignment
issues that often make the SUSY models based on continuous flavour
symmetries rather cumbersome due a proliferation of extra degrees of
freedom.

\subsection{The field content and symmetry breaking}

The full set of the effective theory matter, Higgs and flavon fields
and their transformation properties are given in Table
\ref{tab-fields}. We embed the left-handed Standard Model matter
fields into a triplet of $A_4$ while keeping the right-handed matter
transform as the $SO(3)$-like $A_4$ singlet\footnote{There are in
general three inequivalent one-dimensional representations of $A_4$;
our choice follows the observation that at the string level the extra
singlets (i.e. $SO(3)$ non-invariant ones) usually come from higher
representations of the gauge group and thus seem disfavoured, at least
in simplest schemes.}.  Apart of the pair of MSSM light Higgs doublets
$h$ (arranged into the traditional Pati-Salam bidoublet) driving the
electroweak symmetry breakdown we use a pair of heavy Higgs bosons $H$
and $H'$ to break the Pati-Salam gauge symmetry at a high scale and provide the Majorana mass terms for the right-handed neutrinos.
\TABLE[ht]{
\begin{tabular}{|c|c|c|c|c|}
\hline
field & $SU(4)\otimes SU(2)_{L}\otimes SU(2)_{R}$ & $A_4$ & $U(1)$  & $Z_{2}$\\
\hline
${F}$ &  $(4,2,1)$ & $3$ & 0 & $+$\\
$F^{c}_{1}$   & $(\overline{4},1,2)$ & $1$ & $+2$ & $-$ \\
$F^{c}_{2}$  & $(\overline{4},1,2)$ & $1$ & $+1$ & $+$\\
$F^{c}_{3}$   & $(\overline{4},1,2)$ & $1$ & $-3$ & $-$\\
\hline
$h$  & $(1,2,2)$  &$1$ & 0 & $+$\\
$H$, $\overline{H}$  & $(4,1,2)$, $(\overline{4},1,2)$ & $1$ &
$\pm 3$ & $+$
\\
$H'$, $\overline{H'}$   & $({4},1,2)$,
$(\overline{4},1,2)$ & $1$ & $\mp 3$ & $+$\\
$\Sigma$  & $(15,1,3)$  &$ 1 $ & -1 & $-$\\
\hline
${\phi}_{1}$  & $(1,1,1)$ & $3$ & $+4$ & $+$ \\
${\phi}_{2}$  & $(1,1,1)$ & $3$ & $0$ & $+$\\
${\phi}_{3}$  & $(1,1,1)$ & $3$ & $+3$ & $-$ \\
${\phi}_{23}$  & $(1,1,1)$ & $3$ & $-2$ & $-$\\
${\tilde{\phi}}_{23}$  & $(1,1,1)$ & $3$ & $0$ & $-$ \\
${\phi}_{123}$  & $(1,1,1)$ & $3$ & $-1$ & $+$\\
\hline
\end{tabular}
\caption{\label{tab-fields}  The basic Higgs, matter and flavon
content of the model. }
}
\subsection{The Yukawa sector}
In what follows we shall use upper indices for $A_4$ triplet
components while the lower indices stand for the various species of
structures in the game.
The symmetries defined above allow for the following contributions to the Yukawa superpotential: 
\bea\label{diracsuperpot1} W_{Y}& = & \frac{1}{M}
y_{23}{F}.{\phi}_{23}F_{1}^{c}h+ \frac{1}{M}
y_{123}{F}.{\phi}_{123}F_{2}^{c}h+ \frac{1}{M}
y_{3}{F}.{\phi}_{3}F_{3}^{c}h +\frac{1}{M^{2}}
y_{GJ}{F}.{\tilde{\phi}}_{23}F_{2}^{c}\Sigma h+
 \\
&+ &\frac{1}{M^{2}} y_{13}{F}. ({\phi}_{2}\times
{{\phi}}_{3}) F_{3}^{c}h+\frac{1}{M^{2}} y'_{13}{F}. ({\phi}_{2}*
{{\phi}}_{3}) F_{3}^{c}h+ \frac{1}{M^{3}}
{y}_{23}^{i}I_{i}({F},{\tilde{\phi}}_{23},{\tilde{\phi}}_{23},{\phi}_{3})
F_{3}^{c}h +\ldots \nn 
\eea 
where $x\times y$ is the standard $SO(3)$
cross-product, $(x*y)^i=s^{ijk}x^{j}y^{k}$ (with $s^{ijk}$ being $+1$
for each permutation of $\{i,j,k\}\in\{1,2,3\}$) corresponds to the
extra (symmetric) vector product in $A_4$ while $I_{i}(x,y,u,v)$
denotes the available independent quartic $A_4$ invariants,
as discussed in Appendix~\ref{AppendixA4}. Note that
$M$ is a generic symbol for the mass of the relevant messenger sector
fields giving rise to the desired effective vertices in
\eq{diracsuperpot1}. For sake of conciseness, we shall not discuss the
messenger sector here and defer an interested reader to the study
\cite{King:2006me} for an example of such analysis.

After the spontaneous breakdown of the flavour symmetry (for details see Section \ref{vacuumalignment} and Table \ref{tab-vacuum}) the Yukawa matrices generated from this superpotential piece read:
\be
\label{diracpart}
Y^{f}_{LR}=\left(\begin{array}{ccc}
0 &y_{123} \varepsilon^f_{123} &  \overline{y}_{13} \varepsilon^f_{2} \varepsilon^f_{3} \\
y_{23} \varepsilon^f_{23} & y_{123} \varepsilon^f_{123}+  C^f y_{GJ}\tilde{\varepsilon}^f_{23}\sigma & \overline{y}_{23}(\tilde{\varepsilon}_{23}^f)^{2} \varepsilon_{3}^f \\
-y_{23}\varepsilon^f_{23} &y_{123} \varepsilon^f_{123} - C^f
y_{GJ}\tilde{\varepsilon}^f_{23}\sigma & y_{3}\varepsilon^f_{3}
\end{array}\right)
\ee
where\footnote{Here $M_{f}$ stands for the relevant messenger mass. Note that as it was pointed out in \cite{King:2006me} the masses of messengers governing the up and down sectors can be very different.} 
$$
\varepsilon_{x}^{f}\equiv \frac{|\langle\phi_{x}\rangle|}{M_{f}},
$$
parametrize the  relevant flavon VEVs normalized to the masses of the corresponding messenger fields, 
$C^f=-2,0,1,3$ (for $f=u,\nu, d,e$) are the
traditional Clebsch-Gordon coefficients entering the effective Yukawa vertex in (\ref{diracpart}) including the Higgs field $\Sigma$ (transforming like $(15,1,3)$ under the Pati-Salam symmetry) responsible for the
distinct charged sector hierarchies \`a la Georgi and Jarlskog \cite{Georgi:1979df} and $\sigma$ denotes the
VEV of the Georgi-Jarlskog field  $\sigma\equiv
\langle\Sigma\rangle/M_f$. The effective couplings $\overline{y}_{23}$ and $\overline{y}_{13}$ stem from the multiple contributions to the 13 and 23 elements of $Y^{f}_{LR}$ due to the higher number of relevant cubic and quartic $A_4$ invariants.
\subsection{The Majorana sector}
The Majorana mass matrix is obtained from the superpotential of the form
\bea\label{Maj} 
W_{M} & = &
\frac{1}{M_{\nu}^{3}}w_{1}{F_{1}^{c}}^{2}H H' \phi_{23}^{2}+
\frac{1}{M_{\nu}^{3}}w_{2}{F_{2}^{c}}^{2}H H' \phi_{123}^{2}+
\frac{1}{M_{\nu}}w_{3}{F_{3}^{c}}^{2}H^{2}+
\nn \\
&+ & 
\frac{1}{M^{4}}{F_{1}^{c}}^{2}{H'}^{2}
\left[
w_{4} (\phi_{123} \times {\tilde\phi}_{23}).\phi_{3}+
w'_{4}(\phi_{123}* {\tilde\phi}_{23}).\phi_{3}
\right]
+ \nn \\
&+&
\frac{1}{M^{4}}{F_{1}^{c}}{F_{2}^{c}} H H' \left[w_{5}(\phi_{23} \times \phi_{123}).{\phi}_{2}+w'_{5}(\phi_{23} * \phi_{123}).{\phi}_{2}\right]+ \nn \\
& + & 
\frac{w^{i}_{6}}{M^{5}}{F_{2}^{c}}^{2}{H'}^{2}
I_{i}(\phi_{3},\phi_{3},\phi_{23},{\tilde\phi}_{23})+ 
\frac{w^{i}_{7}}{M^{5}}{F_{1}^{c}}{F_{2}^{c}} {H'}^{2} I_{i}(\phi_{2},\phi_{3},
{\tilde\phi}_{23},{\tilde\phi}_{23})+\ldots \nn
\eea
where as before $I_{i}$ stand for the various $A_4$ quartic invariants and the ellipsis denotes the higher order terms.
It is easy to verify that the Majorana mass matrix emerging from here reads
 \be\label{majoranapart2}
M^{\nu}_{RR}=
\left(\begin{array}{ccc}
{\cal O}(\varepsilon^{\nu 2}_{23}\delta_{H},\varepsilon^{\nu}_{123}\tilde{\varepsilon}^{\nu}_{23}\varepsilon^{\nu}_{3}\delta_{H}^{2}) & . & . \\
. & {\cal O}(\varepsilon^{\nu 2}_{123}\delta_{H},\varepsilon^{\nu 2}_{3}\tilde{\varepsilon}^{\nu}_{23}{\varepsilon}^{\nu}_{23}\delta_{H}^{2}) & . \\
. & . &{\cal O}(1)
\end{array}\right)
\frac{\vev{H}^{2}}{M} 
\ee
where only the relevant terms are displayed because the mixing in the right-handed neutrino sector due to the off-diagonal terms is negligible.
\subsection{The generic results}
In order to achieve a good fit to all the quark and lepton masses and
mixing parameters one has to assume a hierarchy among the flavon VEV
parameters $\varepsilon_{x}^{f}$. Since the relevant VEV scales 
emerge from a radiative symmetry breaking mechanism, as discussed in Section \ref{vacuumalignment},
it is completely natural to expect a certain hierarchy among them that
in turn propagates to the order of magnitude differences in
$\varepsilon_{x}^{f}$. 
The only extra assumption concerns the magnitude of the VEV of $H'$ entering the Majorana sector analysis $\langle H'\rangle\equiv \delta_{H}\langle H\rangle$ with $\delta_{H}\ll 1$. However, a similar radiative mechanism like in the flavon case can play a role here thus making such an assumption as natural as the previous ones.

As it was shown previously in the context of an
$SO(3)$ model \cite{King:2006me}, the structures under consideration
lead to a good fit of all the quark and lepton mass and mixing data
provided\footnote{Note that the role of the $\phi_{12}$ flavon of
\cite{King:2006me} is played here by $\phi_{2}$ with an
advantage of a particular simplicity of the vacuum alignment
mechanism, see Section \ref{vacuumalignment}.  Moreover, the
difference among the vacuum structure of the flavons associated with
the Georgi-Jarlskog mechanism \cite{Georgi:1979df} (i.e. $\tilde\phi_{23}$) with respect to
\cite{King:2006me} is essentially harmless for the fit of the quark and
lepton masses and also the CKM parameters coming predominantly from
the above-diagonal entries are expected to remain stable enough. Thus,
there is no need to perform a dedicated numerical analysis and the
interested reader is again deferred to the one given in
\cite{King:2006me}.}
$\delta_{H}, \varepsilon_{123,23,2}^{f}/\varepsilon_{3}^{f}\sim {\cal O}(10^{-3})$
while $\tilde{\varepsilon}_{23}^{f}/\varepsilon_{3}^{f}\sim {\cal
O}(1)$:
\begin{itemize}
\item{The naturalness of the hierarchy among the third and second generation Yukawa couplings as well as a moderate suppression of the $V_{cb}$ CKM mixing parameter are traced back to the higher-order origin of the relevant (Georgi-Jarlskog and 2-3 entry) operators.}
\item{The first generation masses as well as the smallness of the $V_{ub}$ CKM mixing descend from the hierarchy of the relevant flavon VEVs as discussed in the next section.}
\item{The neutrino sector conforms to the CSD conditions
\cite{King:2005bj}. The particular structure of the neutrino Yukawa
matrix together with the hierarchy of the charged lepton Yukawa couplings leads
to approximate tri-bimaximal mixing in the neutrino sector
\cite{Harrison:2002er} characterized by the
approximately
maximal atmospheric mixing $\tan \theta_{23}\approx 1$, large solar
mixing angle obeying $\sin \theta_{12}\approx 1/\sqrt{3}$ and a small reactor angle
$\theta_{13}\approx 0$, in good agreement with the latest neutrino
data, see e.g. \cite{Strumia:2005tc,Strumia:2006db} and
references therein.}
\item{Concerning the light neutrino mass spectrum, the large hierarchy in $Y^{\nu}_{LR}$ is effectively undone in the seesaw formula by the particular form of the Majorana mass matrix (\ref{majoranapart2}).}
\end{itemize}
Thus, the model provides a very good description of all the known quark and lepton masses and mixing parameters. The only missing ingredient is the mechanism leading to the desired correlations among the VEVs of the various triplet flavon components shown in Table \ref{tab-vacuum}. 
\subsection{The vacuum alignment mechanism \label{vacuumalignment}}
The discrete nature of the flavour
symmetry leads to a particularly simple option to achieve all the
desired vacuum structures displayed in Table \ref{tab-vacuum}. As
discussed in \cite{deMedeirosVarzielas:2006fc}, in such a class of
models the supergravity
(SUGRA) induced D-terms can naturally lead to a set of extra
quartic terms in the effective scalar potential. Such a set of terms,
however, lead to a lift of the would-be degenerate vacua potentially
emerging in a continuous case and thus makes the vacuum alignment
mechanism straightforward.  To force the system to depart from the
symmetric state we shall assume a variant of a radiative symmetry
breaking mechanism, as we now discuss.

Let us first consider the case of a single triplet $\phi$.  Apart from
the obvious $SO(3)$ invariant $(\phi^{\dagger}\phi)^{2}$ the discrete
$A_4$ symmetry admits for instance a contraction like
\be\label{quartic0} I_{0}(\phi^{\dagger},\phi,
\phi^{\dagger},\phi)\equiv \sum_{i=1}^{3}\phi^{\dagger i}\phi^{i}
\phi^{\dagger i}\phi^{i} \ee that breaks the rotational degeneracy of
the would-be $SO(3)$ symmetric vacua.  Assuming that the scalar
potential is governed by the terms\footnote{Here we choose to write
the standard $SO(3)$-invariant term $\Lambda (\phi^{\dagger}\phi)^{2}$
in the basis that exhibits the convexity of the potential rather than
in terms of the ``$I_{1..7}$'' independent invariant advocated in
Appendix \ref{AppendixA4}. It is indeed trivial to see that
$(\phi^{\dagger}\phi)^{2}=(I_{0}+2I_{1})(\phi,\phi^{\dagger},\phi,\phi^{\dagger})$.}
\be
\label{A4potential} V\ni -M^{2}_{\phi}(\phi^{\dagger}\phi)+\lambda
I_{0}(\phi^{\dagger},\phi, \phi^{\dagger},\phi)+\Lambda
(\phi^{\dagger}\phi)^{2}+\ldots 
\ee 
it is easy to verify that the only
vacuum structures that can arise in such a case (i.e. when all the
mixing terms are negligible) are 
\be
\label{available-vacua} \langle|\phi|\rangle\propto (1,1,1)\quad
{\rm and/or}\quad \langle|\phi|\rangle\propto
(1,0,0),\,(0,1,0),\,(0,0,1) 
\ee 
where only the magnitudes of the components are so far specified.
What matters is the sign of the
$SO(3)$-breaking term $\lambda I_{0}(\phi^{\dagger},\phi,
\phi^{\dagger},\phi)$: if $\lambda > 0$ the ``isotropic'' option
$\langle | \phi |\rangle\propto (1,1,1)$ is picked up while the VEV is
maximally ``anisotropic'' (i.e. with just one nonzero entry in
$\langle\phi\rangle$) if $\lambda <0 $. Let us stress that the configurations (\ref{available-vacua}) correspond to the case of an entirely hermitian field
$\phi$. Since both the $I_{0}$ and $I_{1}$ invariants (c.f. Appendix
\ref{AppendixA4}) dominating the scalar potential (\ref{A4potential})
are phase-blind, the current mechanism does not specify the phase of
any of the triplet components if $\phi\neq \phi^{\dagger}$. 
However, as it was shown in \cite{King:2005bj}, what matters in
achieving tri-bimaximal mixing via CSD is 
not the absolute phases of the flavon components but 
the equality of their magnitudes,
and their complex orthogonality, which we shall shortly discuss.

All this leads us to the following realization  of the vacuum alignment
mechanism: suppose each of the 
fields $\phi_{123}$, $\phi_{1}$ and $\phi_{3}$ 
has a potential of the form in Eq.(\ref{A4potential}),
simply repeated for each field.
Suppose each of the fields develop negative
mass-squares through radiative effects around scales $M_{123}$,
$M_{1}$ and $M_{3}$ and let us arrange the ``$\lambda$-terms'' in the
leading piece of the scalar potential in Eq.(\ref{A4potential})
so that they pick up VEVs in the directions allowed by
\eq{available-vacua}, in particular\footnote{The alignment of
  $\langle|\phi_1|\rangle\propto(1,0,0)$ and
  $\langle|\phi_3|\rangle\propto(0,0,1)$ is a just a choice of
basis that we are free to make as long as there are no interactions
binding the VEVs of $\phi_1$ and $\phi_3$ together. On the other hand,
to make sure $\phi_1$ does not coincide with $\phi_3$ spontaneously a
mixing term like $|\phi_1^{\dagger}.\phi_3|^2$ with a positive coefficient can be exploited.}:
\be\label{first-stage}
\langle|\phi_{123}|\rangle\propto (1,1,1)\quad {\rm and} \quad \langle|\phi_{1}|\rangle\propto (1,0,0),\quad \langle|\phi_{3}|\rangle\propto (0,0,1)
\ee
The stability of such a setup requires that the mixing 
arising from the ``inhomogeneous'' terms like\footnote{Here we again
  suppress all the triplet indices so that the generic symbols
  $I_{i}(\phi^{\dagger}_{A}, \phi_{B}, \phi^{\dagger}_{C}, \phi_{D}) $ account
  for the various linearly independent $A_4$ contractions. There are only 4 such
  independent structures for $A=C$ and $B=D$, three if $A=B=C=D$ and 2 if on top of that $\phi=\phi^{\dagger}$ (i.e. only if $\phi$ is strictly neutral), for more details see Appendix \ref{AppendixA4}.
} $
  I_{i}(\phi_{123}^{\dagger},\phi_{123},\phi_{1,3}^{\dagger},\phi_{1,3})$
  , $i\in\{0,1,3\}$ should  be suppressed with respect to the ``pure'' ones
$
I_{i}(\phi_{123}^{\dagger},\phi_{123},\phi_{123}^{\dagger},\phi_{123})$ 
and 
$I_{i}(\phi_{1,3}^{\dagger},\phi_{1,3},\phi_{1,3}^{\dagger},\phi_{1,3})$.

Subsequently, $\langle\tilde{\phi}_{23}\rangle$ and
$\langle{\phi}_{23}\rangle$ can be generated if the interactions with
the first stage fields $\phi_{123}$ and $\phi_{1,3}$ are dominated by
the terms 
\be\label{SO3potential} V\ni
-{M}_{23}^{2}|{\phi}_{23}|^{2}+\lambda_{123}|\phi_{123}^{\dagger}.{\phi}_{23}|^{2}+\lambda_{1}|\phi_{1}^{\dagger}.{\phi}_{23}|^{2}-\tilde{M}_{23}^{2}|\tilde{\phi}_{23}|^{2}+\tilde{\lambda}_{123}|\phi_{123}^{\dagger}.\tilde{\phi}_{23}|^{2}+\tilde{\lambda}_{1}|\phi_{1}^{\dagger}.\tilde{\phi}_{23}|^{2}+\ldots
\ee 
where the ellipsis stands for $SO(3)$ (and thus also $A_4$)
invariant terms of the form $\Lambda_{\phi} (\phi^{\dagger}.\phi)^{2}$
necessary to lift the flat directions. If $\lambda_{123}$ and
$\tilde{\lambda}_{123}$ are positive, the VEVs of $\phi_{23}$ and
$\tilde{\phi}_{23}$ driven to the directions orthogonal to
$\langle{\phi}_{123}\rangle$ while ${\lambda}_{1},\tilde\lambda_{1}>0$
make their first component vanish and thus
$\langle|{\phi}_{23}|\rangle,\langle|{\tilde\phi}_{23}|\rangle\propto
(0,1,1)$.  Concerning the above mentioned ambiguity in fixing the
phases of the vacuum alignment emerging from the simple potential
(\ref{A4potential}), in particular $\phi_{123}$, the orthogonality
condition
$\langle\phi_{123}\rangle^{\dagger}.\langle\phi_{23}\rangle=0$
together with $\langle\phi_{23}^{1}\rangle= 0$ following from the
minimisation of (\ref{SO3potential}) is just enough to generate
$\theta_{13}^{\nu}$ close to zero \cite{King:2002nf} and
$\tan\theta_{12}^{\nu}\sim 1/\sqrt{2}$ regardless any particular arrangement
of the $\langle\phi_{123}\rangle$ phases.
The minus signs in Table \ref{tab-vacuum} are for illustrative purposes
and simply denote the $\pi$-shift in the relative phases of the components
of $\langle\phi_{123}\rangle$ and $\langle\phi_{23}\rangle$ (or
$\langle\tilde\phi_{23}\rangle)$ arising from the relevant
orthogonality conditions.

At this point it is perhaps worth mentioning that the {\it positivity}
of the $\lambda$-couplings above also ensures a better control over
the magnitudes of the corresponding VEVs unlike the case of having an
interaction with a negative coupling constant when a potentially large
negative correction must be compensated by the explicit mass entry
from the F-terms. This means that one can handle easily all the relevant scales
without a need of an extra tuning of parameters in the would-be
``effective wrong-sign masses'' that might otherwise arise.

Concerning the alignment of $\phi_{2}$, a particular shape of its VEV
is immaterial as long as it admits a nonzero projection to the second
$SO(3)$ coordinate. A particularly elegant setup can be obtained if
for instance $\langle|\phi_{2}|\rangle\propto (0,1,0)$ is generated via the same
mechanism like $\phi_{1,3}$, repeating just as before the form of
potential (\ref{A4potential}) with $\lambda_{2}>0$. To make sure the
$(0,1,0)$ option is picked up one can employ the orthogonality
of all the $\phi_{1}$, $\phi_{2}$ and $\phi_{3}$ VEVs via the mixing
terms of the form $|\phi_{i}^{\dagger}.\phi_{j}|^{2}$ with positive
coefficients so that the complete basis of the triplet space is
spanned.

\TABLE[ht]{
\begin{tabular}{|c|c|c|}
\hline
flavon VEV & VEV direction & VEV normalization (scale) \\
\hline
$\langle{\phi}_{1}\rangle$  & $(1,0,0)$ & ${M_{1}}/{\sqrt{2(\lambda_{1}+\Lambda_{1})}}$   \\
$\langle{\phi}_{2}\rangle$  & $(0,1,0)$ &  ${M_{2}}/{\sqrt{2(\lambda_{2}+\Lambda_{2})}}$   \\
$\langle{\phi}_{3}\rangle$  & $(0,0,1)$ & ${M_{3}}/{\sqrt{2(\lambda_{3}+\Lambda_{3})}}$   \\
$\langle{\phi}_{23}\rangle$  & $(0,1,-1)$ &  ${M_{23}}/{2\sqrt{\Lambda_{23})}}$ \\
$\langle{\tilde{\phi}}_{23}\rangle$  & $(0,1,-1)$ &${\tilde{M}_{23}}/{2\sqrt{\tilde{\Lambda}_{23})}}$  \\
$\langle{\phi}_{123}\rangle$  & $(1,1,1)$ &  ${M_{123}}/{\sqrt{2(\lambda_{123}+3\Lambda_{123})}}$  \\
\hline
\end{tabular}
\caption{\label{tab-vacuum} The vacuum alignment pattern generated by the
mechanism specified in the text. The mass scales ${M}_{i}$ and the relevant quartic couplings $\lambda_{i}$ and $\Lambda_{i}$ are defined in Section \ref{vacuumalignment}, Eqs. (\ref{A4potential}) and (\ref{SO3potential}). In the ``VEV direction'' column only the magnitudes of the relevant (in general complex) flavon VEVs are displayed. The minus sign in the case of $\phi_{23}$ and $\tilde\phi_{23}$  illustrates the important $\pi$-difference of the 2nd and 3rd component VEV phases of $\phi_{23}$ and $\phi_{123}$.}  
} 

The results of our vacuum alignment mechanism are sumarized in Table \ref{tab-vacuum}.
It is easy to see that all the mass scales in Table
\ref{tab-vacuum} are essentially free: since the potential (\ref{SO3potential}) is fully $SO(3)$ invariant, the anisotropy enters only through the $A_4$ terms driving the VEVs of $\phi_{123}$ and $\phi_{1,2,3}$ while $\langle\tilde{\phi}_{23}\rangle,\langle{\phi}_{23}\rangle$ can rotate freely to follow the constraints imposed through the interactions with $\phi_{123,1}$. Thus, even a small push in any particular direction is enough to imprint the desired alignment to all the relevant VEVs and we are free to choose $M_{23}$ and $M_{123}$ so that the desired VEV hierarchy is achieved.
\section{Conclusions}
We have constructed the first complete model of flavour based on 
$A_4$ family symmetry together with the
$SU(4)_{C}\otimes SU(2)_{L}\otimes SU(2)_{R}$ Pati-Salam gauge
symmetry. $A_4$ corresponds to the symmetry of the tetrahedron, and is
a discrete subgroup of $SO(3)$. Assuming the simple 
extra symmetry factors $U(1)\otimes Z_{2}$, we have performed
an operator analysis of the model, and shown that the
resulting effective Yukawa and Majorana couplings have a similar
form to those discussed in \cite{King:2006me},
and when the messenger sector is completed, the resulting structures provide a good description of the fermion mass and mixing spectrum.
In particular, the constrained sequential dominance is realized and tri-bimaximal neutrino mixing 
results, with calculable deviations expressed in terms of 
neutrino sum rules \cite{King:2005bj,Antusch:2005kw}.
The main simplification afforded by the discrete symmetry is in 
the vacuum alignment sector. Due to the discrete nature of the
flavour symmetry a particularly simple vacuum mechanism emerges through
the SUGRA induced D-term contributions to the effective scalar
potential that lift the $SO(3)$ vacuum degeneracy. We have shown that this
discrete version of the radiative symmetry breaking mechanism 
may be achieved with a minimal number
of fields that do not participate directly in the Yukawa sector,
and that a realistic model can be constructed which incorporates
all these features simultaneously. The $A_4$ model presented here
may be regarded as being on the same footing as the $\Delta (27)$ model
presented in \cite{deMedeirosVarzielas:2006fc}

\section*{Acknowledgement}
The authors are very grateful to Stefan Antusch for useful comments
and discussions throughout preparing the manuscript.
\appendix
\section{Basic properties of the quartic triplet  $A_4$
  invariants} \label{AppendixA4}
In this appendix we give a short compendium on the main features of
the symmetry group of tetrahedron called $A_4$ and the properties of
the basic triplet invariants used in the main text. Where suitable we use the notation of He et al. \cite{He:2006dk}. 
 
As a discrete subgroup of $SO(3)$,
$A_4$ admits a triplet representation ${\bf 3}$ and three
inequivalent singlets often denoted by ${\bf 1}$, ${\bf 1'}$ and ${\bf
  1''}$. Concerning namely the triplets $x=(x^1,x^2,x^3)$ it is convenient to express
the action of the group  elements by means of its correspondence to the
semidirect product $Z_{3}\ltimes Z_{2}$ as: 
\bea \label{actionA4}
Z_{3}(x^{1},x^{2},x^{3}) & \to & (x^{2},x^{3},x^{1}) \\ 
Z_{2}(x^{1},x^{2},x^{3}) & \to & (x^{1},-x^{2},-x^{3}) \nonumber
\eea     

With this information at hand one can see that apart of the
``standard'' $SO(3)$-like dot product $(x.y)\equiv
x^1y^1+x^2y^2+x^3y^3$ and the cross product $(x\times y).z=\varepsilon^{ijk} x^i
y^j z^k$ there is an extra
symmetric cubic invariant like $(x * y).z=s^{ijk} x^i
y^j z^k$ where $s^{ijk}$ is $+1$ on all permutations  $\{i,j,k\}\in\{1,2,3\}$ and zero otherwise.

At the quartic level one can easily check that the basic structures 
\bea\label{basic}
I_{0}(x,y;u,v) & \equiv &
x_{1}y_{1}u_{1}v_{1}+x_{2}y_{2}u_{2}v_{2}+x_{3}y_{3}u_{3}v_{3} \nonumber \\
I_{1}(x,y;u,v) & \equiv & x_{1}y_{1}u_{2}v_{2}+x_{2}y_{2}u_{3}v_{3}+x_{3}y_{3}u_{1}v_{1} \\
I_{2}(x,y;u,v) & \equiv & x_{1}y_{1}u_{3}v_{3}+x_{2}y_{2}u_{1}v_{1}+x_{3}y_{3}u_{2}v_{2} \nonumber
\eea
are invariant with respect to the action (\ref{actionA4}).However,
this is not the end of the story yet as one must consider also the other permutations of the set of parameters $\{x,y,u,v\}$. The symmetries of $I_{0,1,2}$ are such that all these expressions are actually invariant with respect to permutations of the first and second pair of arguments (and in case of $I_{0}$ even all of them), and thus what matters is just the {\it pairings} of  $\{x,y,u,v\}$. In short, the independent structures emerging from \eq{basic} correspond to $I_{0}$, $I_{1}$ and $I_{2}$ with arguments $(x,y;u,v)$, $(x,u;y,v)$ and $(x,v;y,u)$ only. Due to the maximal symmetry of $I_{0}$ one gets only 4 additional relevant structures from $I_{1}$ and $I_{2}$, namely
\bea
I_{3}(x,y;u,v)  \equiv  I_{1}(x,u;y,v)  &, & 
I_{4}(x,y;u,v)  \equiv  I_{2}(x,u;y,v)  \nonumber \\
I_{5}(x,y;u,v)  \equiv  I_{1}(x,v;y,u) &, & 
I_{6}(x,y;u,v)  \equiv  I_{2}(x,v;y,u) \nonumber
\eea
To demonstrate the completeness of such a ``naively'' constructed set
of invariants it is sufficient to find a mapping of $I_{0,..,6}$ onto
the set of ``group-theoretical'' purely triplet quartic invariants of the form
$(x.y)(u.v)=(3\otimes 3)_{1}\otimes (3\otimes 3)_{1}$,
$(x,y)_{1'}(u,v)_{1''}=(3\otimes 3)_{1'}\otimes (3\otimes 3)_{1''}$ (provided $(x,y)_{1'}\equiv x_{1}y_{1}+\omega x_{2}y_{2}+\omega^{2}x_{3}y_{3}$ and  $(x,y)_{1'}\equiv x_{1}y_{1}+\omega^{2} x_{2}y_{2}+\omega x_{3}y_{3}$ stand for the distinct extra $A(4)$ singlets, $\omega$ is the cubic root of unity and $1'\otimes 1''=1$) ,
$(x\times y).(u\times v)=[(3\otimes 3)_{3a}\otimes (3\otimes 3)_{3a}]_{1}$,
$(x * y).(u\times v) =[(3\otimes 3)_{3s}\otimes (3\otimes 3)_{3a}]_{1}$ and 
$(x * y).(u *  v) = [(3\otimes 3)_{3s}\otimes (3\otimes 3)_{3s}]_{1}$,
leading to a large number of options upon taking into account the various permutations of 4 objects $\{x,y,u,v\}$. However, it is obvious that they are by far not all linearly independent, as one can see for instance from the identity
$
(x\times y).(u\times v)=(x.u)(y.v)-(x.v)(y.u)
$. Indeed, these structures can be mapped onto $I_{0,..,6}$ as 
\bea
2I_{0}(x,y;u,v)  & = & 2(x.y)(u.v)-[(x * v).(y * u)+(x\times v).(y\times u)]
 \\
4I_{1}(x,y;u,v)& = & [(x * v).(y * u)+(x\times v).(y\times u)+(x\times v).(y * u)+(x * v).(y \times u)]
\nonumber \\
4I_{2}(x,y;u,v) & = & [(x * v).(y * u)+(x\times v).(y\times u)-(x\times v).(y * u)-(x * v).(y \times u)]
\nonumber \\
4I_{3}(x,y;u,v) & = & [(x * v).(y * u)-(x\times v).(y\times u)+(x\times
  v).(y * u)-(x * v).(y \times u)]
\nonumber \\
4I_{4}(x,y;u,v) & = & [(x * v).(y * u)-(x\times v).(y\times u)-(x\times
  v).(y * u)+(x * v).(y \times u)]
\nonumber \\
4I_{5}(x,y;u,v) & = & [(x * u).(y * v)-(x\times u).(y \times v)+(x\times u).(y * v)-(x * u).(y \times v)]
\nonumber \\
4I_{6}(x,y;u,v) & = & [(x * u).(y * v)-(x\times u).(y \times v)-(x\times u).(y * v)+(x * u).(y \times v)]\nonumber
\eea
It is then easy to see that whenever $x=u$ and $y=v$ only 4
of these structures remain independent (for instance $I_{0}$, $I_{3}$, $I_{4}$ and one from the equal $I_{1,2,5,6}$). If on top of that $y=x^{\dagger}$ then $I_{3}=I_{4}^{\dagger}$ that allows for only three independent terms in a hermitean scalar potential and, finally, if all the arguments coincide there is
only 2 such terms like for instance $I_0$ and one from $I_{1,..,6}$ left.  
\providecommand{\href}[2]{#2}\begingroup\raggedright\endgroup

\end{document}